\documentclass[11pt]{article}

\usepackage{amssymb,amsthm,amscd,array}
\usepackage{graphics}
\usepackage[final]{epsfig}

\let\ssection=\section
\renewcommand{\section}{\setcounter{equation}{0}\ssection}

\newcommand{\bbR}{\mathbb{R}}
\newcommand{\bbN}{\mathbb{N}}

\newcommand{\bbC}{\mathbb{C}}

\newcommand{\cB}{{\cal{B}}}
\newcommand{\cD}{{\cal{D}}}
\newcommand{\gD}{\mathrm{gr}{\cal{D}}}

\newcommand{\cK}{{\cal{K}}}
\newcommand{\cL}{{\cal{L}}}
\newcommand{\cS}{{\cal{S}}}

\newcommand{\cF}{{\cal{F}}}
\newcommand{\Hom}{\mathrm{Hom}}

\newcommand{\Id}{\mathrm{Id}}

\newcommand{\Sl}{\mathrm{sl}}

\newcommand{\osp}{\mathrm{osp}}

\newcommand{\Supp}{\mathrm{Supp}}

\newcommand{\Vect}{\mathrm{Vect}}

\newcommand{\half}{\frac{1}{2}}
\newcommand{\fg}{\mathfrak{g}}

\chardef\s=110
\chardef\g=103

\begin{document}

\newtheorem{thm}{Theorem}[section]
\newtheorem{lem}[thm]{Lemma}
\newtheorem{cor}[thm]{Corollary}
\newtheorem{conj}[thm]{Conjecture}
\newtheorem{prop}[thm]{Proposition}
\newtheorem{rmk}[thm]{Remark}
\newtheorem{exe}[thm]{Example}
\newtheorem{defi}[thm]{Definition}

\def\a{\alpha}
\def\b{\beta}
\def\d{\delta}
\def\g{\gamma}
\def\om{\omega}
\def\r{\rho}
\def\s{\sigma}
\def\vfi{\varphi}
\def\vr{\varrho}
\def\l{\lambda}
\def\m{\mu}

\title{Differential operators on supercircle:\\
conformally equivariant quantization and symbol calculus}

\author{H. Gargoubi
\thanks{
I.P.E.I.T., 2 Rue Jawaher Lel Nehru, Monfleury 1008
 Tunis,
Tunisie; hichem.gargoubi@ipeit.rnu.tn,}
\and
N. Mellouli
\thanks{
Universit\'e de Lyon, Lyon, F-69003, France; Universit\'e Lyon 1,
Institut Camille Jordan, Villeurbanne Cedex, F-69622, France; CNRS,
UMR5208, Villeurbanne, F-69622, France;
mellouli@math.univ-lyon1.fr,} \and V. Ovsienko
\thanks{
Universit\'e de Lyon, Lyon, F-69003, France;
Universit\'e Lyon 1,
Institut Camille Jordan,
Villeurbanne Cedex, F-69622,
France; CNRS, UMR5208,
Villeurbanne, F-69622, France;
ovsienko@math.univ-lyon1.fr
}}

\date{}

\maketitle

\begin{abstract}
We consider the supercircle $S^{1|1}$ equipped with the standard
contact structure. The conformal Lie superalgebra $\cK(1)$ acts on
$S^{1|1}$ as the Lie superalgebra of contact vector fields; it
contains the M\"obius superalgebra $\osp(1|2)$.  We study the
space of linear differential operators on weighted densities as a
module over $\osp(1|2)$. We introduce the canonical isomorphism
between this space and the corresponding space of symbols and find
interesting resonant cases where such an isomorphism does not exist.
\end{abstract}

\thispagestyle{empty}

\section{Introduction}

Conformally and projectively equivariant symbol calculus and
quantization (see, e.g., \cite{LO,DLO}) is the canonical way to
define a ``total symbol'' of a differential operator on a manifold
equipped with a $G$-structure (e.g., projective or conformal
structure). In the case of supermanifolds, the only information
available at the moment is for the supercircle $S^{1|1}$ with a
contact structure, see \cite{CMZ}.

The Lie superalgebra, $\cK(1)$ of contact vector fields on
$S^{1|1}$ is often called a {\it superconformal algebra}, cf.
\cite{GLS1}. This algebra and its central extension are the
simplest super-generalizations of the Witt and Virasoro algebras,
respectively. The Lie superalgebra $\osp(1|2)$ plays on $S^{1|1}$
the same crucial role that $\Sl(2)$ plays on $S^{1}$.

The main object of our study is the space of linear differential
operators acting on weighted densities. The space $\cF_\l$ of
weighted densities with weight $\l\in\bbC$ (or $\l$-densities for
short) is a module over $\cK(1)$. Therefore, the space
$\cD_{\l,\m}$ of linear differential operators from $\cF_\l$ to
$\cF_\m$ is also a $\cK(1)$-module. Each module $\cD_{\l,\m}$ has
a natural filtration by the order of differential operators; the
graded module $\gD_{\l,\m}$ is called the {\it space of symbols}.
We restrict the $\cK(1)$-module structures to a particular
subalgebra $\osp(1|2)$ and look for $\osp(1|2)$-isomorphisms
$$
\s:\cD_{\l,\m}\to\gD_{\l,\m},
\qquad
Q:\gD_{\l,\m}\to\cD_{\l,\m},
$$
where $Q=\s^{-1}$. These isomorphisms are called the $\osp(1|2)$-equivariant
\textit{symbol map} and \textit{quantization map}, respectively.

For almost all values $(\l,\m)$, we prove the existence and uniqueness (up to
normalization) of the $\osp(1|2)$-equivariant symbol map $\s$ and calculate
its explicit formula. We also calculate its inverse $Q=\s^{-1}$ generalizing
the quantization map from \cite{CMZ}.

We consider particular values of the weights that we call \textit{resonant}
which satisfy
\begin{equation}
\label{ResValSh}
\m-\l=\frac{1}{2},1,\frac{3}{2},2,\dots
\end{equation}
We prove that for these values there is no $\osp(1|2)$-isomorphism between
the space of linear differential operators and the corresponding space of
symbols, except for the special values:
\begin{equation}
\label{TheResVal}
\l=\frac{1-m}{4},\qquad
\m=\frac{1+m}{4},
\end{equation}
where $m$ is odd. We calculate the cohomological obstructions
to existence of such an isomorphism.

The $\cK(1)$-modules $\cD_{\l,\m}$, where $(\l,\m)$ as in (\ref{TheResVal}),
are of particular interest. These modules are characterized by the existence
of $\osp(1|2)$-invariants. Let us mention that in the classical case
of $S^1$ the corresponding modules are closely related to the
Virasoro and Adler-Gelfand-Dickey algebras, see, e.g., \cite{OT}.

Our main tool is the \textit{finer filtration} on the space of differential
operators $\cD_{\l,\m}$:
\begin{equation}
\label{FiltrForm}
\cD^0_{\l,\m}\subset
\cD^{\frac{1}{2}}_{\l,\m}
\subset\cD^1_{\l,\m}
\subset
\cD^{\frac{3}{2}}_{\l,\m}
\subset\cdots\subset
\cD^{\ell-\frac{1}{2}}_{\l,\m}\subset
\cD^\ell_{\l,\m}
\subset\cdots
\end{equation}
We introduce the notion of differential operators of
{\it semi-integer order}. In particular, the space
$\Vect_\bbC(S^{1|1})$ of all vector fields (\ref{vectfields}) is,
as $\cK(1)$-module, a direct sum of two submodules: $\cK(1)$ itself
and the space of tangent vector fields. The finer filtration
is stable with respect to the $\cK(1)$-action.

It worth noticing that the results of this paper remain true in the case of
pseudodifferential operators, as considered in \cite{CMZ},
but we will not dwell on it.

\section{Geometry of the supercircle}

The \textit{supercircle} $S^{1|1}$ is the simplest supermanifold
of dimension $1|1$ generalizing $S^1$. In order to fix notation,
let us give here the basic definitions of geometric objects on
$S^{1|1}$;  for more details, see \cite{Ber,Lei,LR,GLS}.

We define the supercircle $S^{1|1}$ in terms of its superalgebra
of functions, denoted by $C_\bbC^\infty(S^ {1|1})$ and consisting
of elements of the form
\begin{equation}
\label{SoF}
f(x,\xi)=f_0(x)+\xi\,f_1(x)
\end{equation}
where $x$ is an arbitrary parameter on $S^1$ and $\xi$ is an
(formal Grassmann) coordinate such that $\xi^2=0$.
The parity function $p$ is defined by $p(f(x))=0$ and $p(\xi)=1$.

\subsection{Vector fields and differential forms}

A \textit{vector field} on $S^{1|1}$ is a superderivation of
$C_\bbC^\infty(S^{1|1})$.
Every vector field can be expressed in coordinates in terms of
partial derivatives:
\begin{equation}
\label{vectfields}
 X=f\,\frac{\partial}{\partial{}x}+
g\,\frac{\partial}{\partial\xi},
\end{equation}
where $f,g\in{}C_\bbC^\infty(S^{1|1})$. The space of vector fields is a
Lie superalgebra denoted by $\Vect_\bbC(S^{1|1})$. Let
$\Omega^1(S^{1|1})$ be the rank $1|1$ right
$C_\bbC^\infty(S^{1|1})$-module with basis $dx$ and $d\xi$; we interpret
it as the \textit{right} dual over $C_\bbC^\infty(S^{1|1})$ to the
left $C_\bbC^\infty(S^{1|1})$-module $\Vect_\bbC(S^{1|1})$, by
setting $\langle\partial_{y_i}, dy_j\rangle =\delta _{ij}$ for
$y=(x, \xi)$. The space $\Omega^1(S^{1|1})$ is a {\it left} module
over $\Vect_\bbC(S^{1|1})$, the action being given by the
\textit{Lie derivative}:
$$
\langle X, L_Y\a\rangle:= \langle[X,Y], \a\rangle.
$$

\subsection{Lie superalgebra of contact vector fields}

The standard \textit{contact structure}\footnote{This structure is
famous in mathematical physics, it is also known as the
``SUSY-structure''.} on $S^{1|1}$ is defined as a codimension $1$
{\it non-integrable} distribution $\langle\overline{D}\rangle$ on
$S^{1|1}$, i.e., a subbundle in $TS^{1|1}$ generated by the odd
vector field
\begin{equation}
\label{VFD}
\overline{D}=\frac{\partial}{\partial\xi}-\xi\,\frac{\partial}{\partial{}x}.
\end{equation}
This contact structure can be
equivalently defined as the kernel of the differential 1-form
\begin{equation}
\label{CoFor} \a=dx+\xi\,d\xi.
\end{equation}
A vector field $X$ on $S^{1|1}$ is said to be \textit{contact} if
it preserves the contact distribution:
$$
[X,\overline{D}]=\psi_X\overline{D},
$$
where $\psi_X\in{}C_\bbC^\infty(S^{1|1})$ is a function depending
on $X$. The space of contact vector fields is a Lie superalgebra
denoted by $\cK(1)$. The following statement is well-known.

\begin{lem}
\label{ProContLem}  Every contact vector field can be expressed,
for any $f\in{}C_\bbC^\infty(S^{1|1})$, as\footnote{For
interpretation of the fields $D$ and $\overline{D}$, see
\cite{Shch}.}
\begin{equation}
\label{CvF} X_f=-f\,\overline{D}^2+ \half{}D(f)\,\overline{D},
\quad
\textrm{ where }
\quad
D=\frac{\partial}{\partial\xi
}+\xi\,\frac{\partial}{\partial{}x}.
\end{equation}
\end{lem}

The vector field (\ref{CvF}) is said to be the contact vector
field with \textit{contact Hamiltonian} $f$. One checks that
\begin{equation}
\label{ActAlp}
L_{X_f}\a=f'\,\a,
\qquad
[X_f,\overline{D}]=-\half\,f'\,\overline{D}.
\end{equation}

The \textit{contact bracket} is defined by
$
[X_f,X_g]=X_{\{f,g\}}.
$
The space $C_\bbC^\infty(S^{1|1})$ is thus equipped with a Lie superalgebra
structure isomorphic to $\cK(1)$.
The explicit formula can be easily calculated:
\begin{equation}
\label{CoBrExp}
\{f,g\}=fg'-f'g+(-1)^{p(f)(p(g)+1)}\,\frac{1}{2}\,D(f)D(g).
\end{equation}

\subsection{Projective/conformal symmetries: $\osp(1|2)$-action}

In the case of $S^1$, the notions of \textit{projective}
and \textit{conformal} structures coincide
and are defined by the action of $\Sl(2)$.
In the adapted (local) coordinate $x$ on $S^1$
this action is spanned by three vector fields:
\begin{equation}
\label{ConProCircForm} \Sl(2)=\mathrm{Span}\left(
\frac{\partial}{\partial{}x},\quad
x\frac{\partial}{\partial{}x},\quad
x^2\frac{\partial}{\partial{}x} \right)
\end{equation}
corresponding to the fraction-linear transformations
$$
x\mapsto
\frac{ax+b}{cx+d},
\qquad
ad-bc=1.
$$
A \textit{projective structure} on $S^1$ is given by an atlas with
fraction-linear coordinate transformations (in other words, by an
atlas such that the $\Sl(2)$-action (\ref{ConProCircForm}) is
well-defined). Classification of projective structures on $S^1$ is
equivalent to classification of coadjoint orbits of the Virasoro
algebra (see \cite{KIR2}, \cite{OT}).

A projective/conformal structure on $S^{1|1}$ is defined as a
(local) action of the orthosymplectic Lie superalgebra $\osp(1|2)$
generated by $D$ and $xD$:
\begin{equation}
\label{ConProSupCircForm}
\begin{array}{l}
\displaystyle \osp(1|2)_0=\mathrm{Span}\left(
X_1=\frac{\partial}{\partial{}x}, \quad
X_x=x\frac{\partial}{\partial{}x}+\half\,\xi\,\frac{\partial}{\partial\xi},
\quad
X_{x^2}=x^2\frac{\partial}{\partial{}x}+x\xi\,\frac{\partial}{\partial\xi}
\right),
\\[14pt]
\displaystyle \osp(1|2)_1=\mathrm{Span}\left( X_{\xi}=D, \qquad
X_{\xi x}=x\,D \right).
\end{array}
\end{equation}

\begin{rmk}
{\rm
The $\osp(1|2)$-action (\ref{ConProSupCircForm}) is the infinitesimal
version of the contact fractional-linear transformations
\begin{equation}
\label{SuperLiFracForm}
(x,\xi)\mapsto
\left(
\frac{ax+b+\g\xi}{cx+d+\d\xi},\,
\frac{\a{}x+\b+e\xi}{cx+d+\d\xi}
\right),
\end{equation}
where
$
ad-bc-\a\b=1,
e^2+2\g\d=1,
\a{}e=a\d-c\g
$
and
$
\b{}e=b\d-d\g
$
(cf. \cite{CMZ}).
}
\end{rmk}

As in the $S^1$ case, these contact fraction-linear
transformations preserve the action (\ref{ConProSupCircForm}) so
that atlases compatible with the projective structure are
precisely those atlases for which the action
(\ref{ConProSupCircForm}) is well defined. For a classification of
projective/conformal structures on $S^{1|1}$, also equivalent to
classification of the orbits in the coadjoint representations of
the Neveu-Schwarz and Ramond superalgebras, see \cite{OOC}.

\begin{rmk}
{\rm Eq. (\ref{ActAlp}) implies that the (super)centralizer of $\overline{D}$
is spanned by $X_\xi=D$ and $X_1=D^2$.}
\end{rmk}

\section{Modules of weighted densities}

We introduce a 1-parameter family of modules over the Lie superalgebra
$\cK(1)$. As vector spaces all these modules are isomorphic to
$C_\bbC^\infty(S^{1|1})$, but not as $\cK(1)$-modules.

For every contact vector field $X_f$, define a 1-parameter family of
first-order differential operators on $C_\bbC^\infty(S^{1|1})$:
\begin{equation}
\label{LieDerLam}
L_{X_f}^\l=
X_f+\l{}f',
\qquad
\l\in\bbC.
\end{equation}
One easily checks the the map $X_f\mapsto{}L^\l_{X_f}$ is a
homomorphism of Lie superalgebras, i.e.,
$[L^\l_{X_f},L^\l_{X_g}]=L^\l_{[X_f,X_g]}$, for every $\l$. One thus
obtains a 1-parameter family of $\cK(1)$-modules on
$C_\bbC^\infty(S^{1|1})$ that we denote $\cF_\l$ and call the space
of \textit{densities of weight} $\l$ (or $\l$-densities for short).
The space of 1-forms proportional to $\a$ is the space of sections
of the line bundle
${\langle\overline{D}\rangle}^\bot\subset{}T_\bbC^*S^{1|1}$ over
$S^{1|1}$, i.e., the line bundle of covectors orthogonal to the
contact distribution. A $\l$-density is a section of the line
bundle
$\left({\langle\overline{D}\rangle}^\bot\right)^{\otimes\l}$. It
is then natural to express every $\l$-density in terms of the
contact form $\a$ as
$$
\phi=f\,\a^\l{},
\quad
\textrm{where}
\quad
f\in{}C^\infty(S^{1|1}).
$$
\begin{exe}
\label{FirstLem}
{\rm
(a)
The module $\cF_0$ is nothing but the space of functions $C^\infty(S^{1|1})$.

(b)
The module $\cF_1$ is the space of 1-forms proportional to $\a$.

(c) A more interesting example is provided by the Lie superalgebra
$\cK(1)\cong\cF_{-1}$ viewed as a module over itself, see below.}
\end{exe}

As $\cK(1)$-module, the space of volume forms on $S^{1|1}$ is
isomorphic (up to parity, perhaps) to $\cF_{\frac{1}{2}}$.
Therefore, \textit{Berezin integral} (\cite{Ber,Lei})
$\cB:\cF_{\frac{1}{2}}\to\bbC$ can be given, for any
$f=f_0(x)+\xi\,f_1(x)$, by the formula
$$
\cB (f \a^\frac{1}{2}):= \int_{S^1}f_1(x)\,dx.
$$
So the product of densities composed with $\cB$ yields a bilinear
$\cK(1)$-invariant form:
\begin{equation}
\label{BilinFor}
\langle\cdot ,\cdot\rangle:
\cF_\l\otimes\cF_{\m}\to\bbC, \qquad \l+\m=\frac{1}{2}.
\end{equation}

It turns out that the adjoint $\cK(1)$-module is isomorphic to
$\cF_{-1}$. In other words, a contact Hamiltonian is a density of
weight $-1$ rather than a function:
\begin{equation}
\label{LieDerPB}
\{f,g\}=L^{-1}_{X_f}\,g.
\end{equation}
This statement can be reformulated as follows.
\begin{cor}
\label{ConBraProCor} For a contact vector field $X_f$ given by
(\ref{CvF}), the expression
\begin{equation}
\label{ConHamDenFor}
\s(X_f):=f\,\a^{-1}
\end{equation}
is a well defined $-1$-density. The expression
(\ref{ConHamDenFor}) is $\cK(1)$-invariant and independent of the
choice of the contact form ~$\a$.
\end{cor}

\subsection{Poisson algebra of weighted densities}

The contact bracket (\ref{CoBrExp}) extends to densities of
arbitrary weight: $\{\,,\,\}:\cF_\l\otimes\cF_\mu\to\cF_{\l+\mu+1}$
and defines a structure of Poisson Lie superalgebra. Explicitly (\cite{LKV})
\begin{equation}
\label{LieDerGeneral}
\{f,g\}=\l{}fg'-\m{}f'g+(-1)^{p(f)(p(g)+1)}\,\frac{1}{2}\,D(f)D(g).
\end{equation}
The following statement can be checked directly (cf. Grozman's
list of invariant operators \cite{Groz}).

\begin{prop}
The operation (\ref{LieDerGeneral}) is $\cK(1)$-invariant and
satisfies the Jacobi and Leibniz identities
makes the space of
weighted densities on $S^{1|1}$ a Poisson superalgebra.
\end{prop}

\subsection{Splitting of vector fields}

\begin{prop}
\label{FullVFPro}
There is an isomorphism of $\cK(1)$-modules
\begin{equation}
\label{fullLAVF}
\Vect(S^{1|1})\cong\cF_{-\frac{1}{2}}\oplus\cF_{-1}.
\end{equation}
\end{prop}
\begin{proof}
The submodule $\cF_{-1}\subset\Vect_\bbC(S^{1|1})$
is the subalgebra $\cK(1)$ itself
(see Corollary \ref{ConBraProCor});
the submodule $\cF_{-\frac{1}{2}}\subset\Vect_\bbC(S^{1|1})$, consists of the
vector fields tangent to the contact distribution:
$X=g\,\overline{D}$, where $g$ is an
arbitrary function.
\end{proof}

Note that an analog of Proposition \ref{FullVFPro} holds for any
contact (super)manifold (see \cite{Ovs}).

\section{Differential operators on weighted densities}

For differential operators on supermanifolds, see \cite{BL, LR, CMZ,BO}.

\subsection{The modules $\cD_{\l,\m}$: definition}\label{DDOSect}

The classical Peetre theorem \cite{Pee} implies that the following
two definitions are equivalent. They reflect two different aspects
of the notion: geometric and algebraic.

1) A linear map $ A:\cF_\l\to\cF_\m $ is called a
\textit{differential operator} if it is \textit{local}, that is,
if it preserves the supports of the arguments:
$\Supp{A(\phi)}\subset\Supp{\phi}$, where the support of a
$\l$-density is a subset of $S^1$ given by the union of the
supports of the even and the odd parts.

2)  A linear map $A:\cF_\l\to\cF_\m$ is a \textit{differential
operator} if there exists an integer $k$ such that
$[\cdots[A,f_1],\cdots,f_{k+1}]=0$ for any functions $f_1,\ldots,f_{k+1}$.
The minimal such $k$ is called the \textit{order} of $A$.

We denote $\cD_{\l,\m}$ the space of linear differential operators
from $\cF_\l$ to $\cF_\m$ and $\cD^k_{\l,\m}$ the space of linear
differential operators of order $k$. One has a filtration
$$
\cD^0_{\l,\m}\subset\cD^1_{\l,\m}
\subset\cdots\subset
\cD^k_{\l,\m}
\subset\cdots
$$

\begin{exe}
{\rm
The space of zeroth-order operators is $\cD^0_{\l,\m}\cong\cF_{\m-\l}$;
it consists of the operators of multiplication by $(\m-\l)$-densities.
The space of first-order operators
on densities of fixed weight $\l=\m$ is
$\cD^1_{\l,\l}\cong\Vect_\bbC(S^{1|1})\oplus{}C_\bbC^\infty(S^{1|1}).$
}
\end{exe}

The space $\cD_{\l,\m}$ is naturally a module over $\cK(1)$;
the action being given by the commutator with Lie derivative:
\begin{equation}
\label{DiffOpKActForm}
\cL^{\l,\m}_{X_f}(A)
:=
L^\m_{X_f}\circ{}A-A\circ{}L^\l_{X_f}.
\end{equation}
The above filtration is obviously $\cK(1)$-invariant.

\begin{prop}
\label{EcritProp} (i) Every differential operator
$A\in\cD_{\l,\m}$ can be expressed in the form:
\begin{equation}
\label{DiffOpGenForm}
A(f\,\a^\l) = \sum_{i=0}^{\ell} a_i(x,\xi)\overline{D}^i\left(f\right)\a^\m,
\end{equation}
where the coefficients $a_i(x,\xi)$ are arbitrary functions and $\ell\in\bbN$.

(ii)
If $A\in\cD^k_{\l,\m}$, then $\ell=2k$.
\end{prop}

\begin{proof}
Part (i). Every DO on $S^1$ is a (finite) expression $ A =
\mathop{\sum}\limits_{i\geq0}
a_i(x)\left(\frac{\partial}{\partial{}x}\right)^i$,  see
\cite{Pee}. It follows that a differential operator on $S^{1|1}$
can be expressed as a $2\times2$-matrix of differential operators
on $S^1$, or, equivalently,
$$
 A = \mathop{\sum}\limits_{i\geq0}
\widetilde{a}_i(x,\xi)\left(\frac{\partial}{\partial{}x}\right)^i+
 \mathop{\sum}\limits_{i\geq0} \widetilde{b}_i(x,\xi)
\left(\frac{\partial}{\partial{}x}\right)^i\frac{\partial}{\partial\xi}.
$$
Since $\frac{\partial}{\partial{}x}=-\overline{D}^2$, every
differential operator $A$ is a polynomial expression in $\overline{D}$.

Part (ii) is straightforward.
\end{proof}

\begin{rmk}
\label{RemOrd}
{\rm
(a)
There is another way to express a differential operator in local coordinates:
\begin{equation}
\label{DiffOpGenFormEquiv}
A(f\,\a^\l) = \sum_{i\geq0} b_i(x,\xi)D^i\left(f\right)\a^\m.
\end{equation}
where the coefficients $b_i$ are related to $a_i$  via
\begin{equation}
\label{AandB}
b_{2i}=(-1)^i(a_{2i}-2\xi\,a_{2i-1}),
\qquad  b_{2i+1}=(-1)^ia_{2i+1}.
\end{equation}
Formula (\ref{AandB}) follows from the expression $\overline{D}=D-2\xi\,D^2$.

(b)
Clearly, $ b(x,\xi)\,D^i \in\cD_{\l,\m}^{\left[\frac{i+1}{2}\right]}, $
where $[\ell]$ is the integral part of a real number $\ell$. }
\end{rmk}

The finer filtration (\ref{FiltrForm}) on the space $\cD_{\l,\m}$
(clearly stable under the $\cK(1)$-action) is given by the spaces
$\cD^{\frac{\ell}{2}}_{\l,\m}$ of differential operators
(\ref{DiffOpGenForm}).

\subsection{Conjugation of differential operators}

There exists a $\cK(1)$-invariant  \textit{conjugation} map
$*:\cD_{\l,\m}\to\cD_{\frac{1}{2}-\m,\frac{1}{2}-\l}$ defined by
\begin{equation}
\label{ConjFor} \langle{}A\phi,\psi\rangle =(-1)^{p(A)p(\phi)}\,
\langle{}\phi,A^*\psi\rangle
\end{equation}
for any $A\in\cD_{\l,\m}$ and
$\phi\in\cF_\l,\,\psi\in\cF_{\frac{1}{2}-\m}$, where
$\langle\cdot,\cdot\rangle$ is the bilinear form (\ref{BilinFor}).

Clearly, $*$ is a $\cK(1)$-isomorphism
$\cD^\ell_{\l,\m}\cong\cD^\ell_{\frac{1}{2}-\m,\frac{1}{2}-\l}$
for every $\ell\in\frac{1}{2}\,\bbN$. In the particular case
$\l+\m=\frac{1}{2}$, the map $*$ is an involution. The module
$\cD^\ell_{\l,\frac{1}{2}-\l}$ splits into a direct sum of the
submodules of \textit{symmetric} and \textit{skew-symmetric} operators.

The explicit formula of the conjugation map is easy to calculate:
\begin{equation}
\label{ConjForExp}
*:\overline{D}^k\mapsto
(-1)^{\left[\frac{k+1}{2}\right]}\,
\overline{D}^k.
\end{equation}

\subsection{The principal symbol map}

The highest order coefficient in (\ref{DiffOpGenForm}) has the
following geometric meaning.

\begin{prop}
\label{QuotPro} For every $\ell\in\frac{1}{2}\,\bbN$, we have
\begin{equation}
\label{DiffOpSplitForm}
\cD^\ell_{\l,\m}/\cD^{\ell-\frac{1}{2}}_{\l,\m} \cong
\cF_{\m-\l-\ell}.
\end{equation}
\end{prop}
\begin{proof}
For a differential operator $A\in\cD^\ell_{\l,\m}$ given by
(\ref{DiffOpGenForm}), one easily checks that the expression
\begin{equation}
\label{PrinSymForm}
\s_{\rm pr}(A):=a_\ell(x,\xi)\,\a^{\m-\l-\ell}
\end{equation}
is a well-defined $(\m-\l-\ell)$-density.
\end{proof}

The $\cK(1)$-invariant projection
\begin{equation}
\label{PrinSymUniForm}
\s_{\rm pr}:\cD^\ell_{\l,\m}\to\cF_{\m-\l-\ell}
\end{equation}
will be called the \textit{principal symbol map}.

\subsection{Space of symbols of differential operators}

Consider the graded $\cK(1)$-module $\gD_{\l,\m}$ associated with the  
filtration
(\ref {FiltrForm}). Proposition \ref{QuotPro} implies that this
$\cK(1)$-module is a direct sum of density modules:
$$
\gD_{\l,\m} = \bigoplus_{i=0}^\infty \cF_{\m-\l-\frac{i}{2}}.
$$
Note that this module depends only on the shift, $\m-\l$, of the
weights and not on $\m$ and $\l$ independently. We call this
$\cK(1)$-module the \textit{space of symbols of differential operators}
and denote it $\cS_{\m-\l}$. The space of $\ell$-th order symbols is
\begin{equation}
\label{SpaceSymForm}
\cS^\ell_{\m-\l} := \bigoplus_{i=0}^{2\ell}
\cF_{\m-\l-\frac{i}{2}},
\quad
\textrm{where}
\quad
\ell\in\frac{1}{2}\,\bbN.
\end{equation}

\section{Non-trivial cohomology classes of Lie superalgebra $\osp(1|2)$}

Given a Lie (super)algebra and its module, what is the
corresponding cohomology ring?
We give here some partial information about
$H^0(\osp(1|2);\cD_{\l,\m})$ (the space of invariants) and
$H^1(\osp(1|2);\cD_{\l,\m})$. More precisely, we exhibit
non-trivial cocyles and conjecture that these cocyles generate the
respective cohomology spaces.

\begin{prop}
\label{BolPro}
For every $k=1,3,5\ldots$, the differential operator
\begin{equation}
\label{BolOpr}
\overline{D}^k:\cF_\frac{1-k}{4}\to\cF_\frac{1+k}{4}
\end{equation}
is $\osp(1|2)$-invariant.
\end{prop}

The operators (\ref{BolOpr}) were found in
\cite{Gie,Gie2}, as analogs of the Bol operators on
$S^1$, (see, e.g., \cite{OT}).
One can prove that Proposition \ref{BolPro}
provides the complete list of $\osp(1|2)$-invariant operators on
weighted densities, but we will not need this.

One more special property of the modules
$\cD_{\l,\m}$ with $(\l,\m)$ given by (\ref{TheResVal}) is as follows.

\begin{thm}
\label{CohomThm}
For every $k=1,3,5\ldots$, the linear map
$
\g_k:\osp(1|2)\to\cD_{\frac{1-k}{4},\frac{1+k}{4}},
$
defined by
\begin{equation}
\label{CocGamma}
\g_k(X_f)=
D^3(f)\,\overline{D}^{k-1}+
\frac{k-1}{2}\,
D^4(f)\,\overline{D}^{k-2},
\end{equation}
where $f$ is the contact Hamiltonian of an element $X_f\in\osp(1|2)$,
is a non-trivial 1-cocycle.
\end{thm}

\begin{proof}
To prove that $\g_k$ is a 1-cocycle, consider a map
$\g^\l_k:\osp(1|2)\to\cD_{\l,\l+\frac{k}{2}}$ defined by the
same formula (\ref{CocGamma}) for arbitrary $\l$. One checks that
$$
L^{\l+\frac{k}{2}}_{X_f}\circ\overline{D}^k
-(-1)^{p(f)}\,\overline{D}^k\circ{}L^\l_{X_f} =
\textstyle \left(\l+\frac{k-1}{4}\right)\g_k(X_f).
$$
Hence,
$\g^\l_k=\frac{4}{4\l+k-1}\,\d\,\overline{D}^k$, so it satisfies
the 1-cocycle condition if $\l\neq\frac{1-k}{4}$.
By continuity, this is also true for $\l=\frac{1-k}{4}$.

The 1-cocycle $\g^\l_k$ is a coboundary for any $\l\neq\frac{1-k}{4}$.
Let us prove that in the case $\l=\frac{1-k}{4}$,
this cocycle is indeed non-trivial.

We will need the explicit formula for $\g_k$ in terms of the basis of
$\osp(1|2)$: it vanishes on all elements $X_f$ of $\osp(1|2)$
except $f=x^2$ or $x\xi$, for which
\begin{equation}
\label{CocGammaEF}
\begin{array}{rcl}
\g_k(X_{x^2})
&=&
2\xi\,\overline{D}^{k-1}+
\frac{k-1}{2}\,
\overline{D}^{k-2},\\[6pt]
\g_k(X_{x\xi})
&=&
\overline{D}^{k-1}.
\end{array}
\end{equation}
Assume that there exists an operator $A\in\cD_{\frac{1-k}{4},\frac{1+k}{4}}$
such that $\g_k$ is equal to $\d\,A$, where
$$
(\d\,A)\,(X_f)=
L^\frac{1+k}{4}_{X_f}\circ{}A
-(-1)^{p(f)}\,A\circ{}L^\frac{1-k}{4}_{X_f}.
$$
The operator $A$ is of the form $A=a_m\,\overline{D}^m+\cdots+a_0$;
its principal symbol is a density of degree $\m-\l-\frac{m}{2}=\frac{k-m}{2}$.
If $m>k$, then, the principal symbol of $(\d\,A)\,(X_f)$ is not
identically zero since there is no $\osp(1|2)$-invariant density (except the
constant function). Therefore, $m\leq{}k$ and
$A=a_k\,\overline{D}^k+\cdots+a_0$, where $a_k=\mathrm{const}$.
According to Proposition \ref{BolPro}, the term $\overline{D}^k$ commutes with
the action of $\osp(1|2)$, so it remains to consider
$A=a_{k-1}\,\overline{D}^{k-1}+\cdots+a_0$ with $a_{k-1}\not\equiv0$.
The principal symbol of $A$ is a $\frac{1}{2}$-density  
$a_{k-1}\,\a^\frac{1}{2}$.
By assumption, the principal symbol of the operator is
$$
\s_{\rm pr}(\g_k(X_f))=
L^\frac{1}{2}_{X_f}
\left(
a_{k-1}\,\a^\frac{1}{2}
\right)
$$
Finally, the relation $\g_k(X_f)=0$ implies that $a_{k-1}$ is
constant contradicting (\ref{CocGammaEF}).
\end{proof}

\begin{rmk}
{\rm
1)
The cocycle $\g_k$ is odd (since $k$ in (\ref{CocGamma}) is odd).

2)
One checks that
$$
\g_k(X_f)^*=(-1)^{\frac{k-1}{2}}\,\g_k(X_f)
\quad
\textrm{for each
$X_f\in\osp(1|2)$. }
$$
}
\end{rmk}

Computation of the cohomology of $\osp(1|2)$ with coefficients in
$\cD_{\l,\m}$ is an interesting open problem. We formulate a conjecture
on the structure of the first cohomology space.

\begin{conj}
One has $H^1(\osp(1|2);\cD_{\l,\m})=\bbC^{0|1}$
if and only if $(\l,\m)=(\frac{1-k}{4},\frac{1+k}{4})$,
with odd $k$ and spanned by the cocycle $\g_k$.
Otherwise, this cohomology space is trivial.
\end{conj}

In the case of $S^1$ and the Lie algebra $\Sl(2)$ a similar result was
obtained in \cite{Lecom}.

\section{Equivariant quantization and symbol maps}

We restrict the $\cK(1)$-action on $\cD^\ell_{\l,\m}$ to the
subalgebra $\osp(1|2)$ and look for an $\osp(1|2)$-isomorphism
between $\cD^\ell_{\l,\m}$ and $\cS^\ell_{\m-\l}$ providing a
``total symbol'' of differential operators. We prove existence and
uniqueness (up to normalization) of such an isomorphism for
generic $(\l,\m)$ and investigate the resonant case.

\subsection{The main theorem}

A map $\s:\cD_{\l,\m}\to\cS_{\m-\l}$ is called a \textit{symbol map} if it is
bijective and for every $\ell\in\frac{1}{2}\,\bbN$ the following diagram
is commutative:
\begin{equation}
\begin{CD}
\label{diag}
\cD^\ell_{\l,\m}@>\s>>\cS^\ell_{\m-\l}\\
@V\s_{\rm pr}VV            @VVV\\
\cF_{\m-\l-\ell}@>\Id>>\cF_{\m-\l-\ell}
\end{CD}
\end{equation}
where the right arrow is the projection. In other words, the highest-order
term of $\s$ coincides with the principal symbol map.
The inverse map, $Q=\s^{-1}$, is called the \textit{quantization map}.

Recall that we call $(\l,\m)$ \textit{non-resonant} if $\m-\l$ is not as in
(\ref{ResValSh}). The following statement is the main result of this paper;
the proof will be given in the next section.

\begin{thm}
\label{MainThm}
(i)
If $\m-\l$ is non-resonant, then $\cD_{\l,\m}\cong\cS_{\m-\l}$ as
$\osp(1|2)$-modules, the isomorphism being given by the unique
$\osp(1|2)$-invariant symbol map
\begin{equation}
\label{SymMapMainFor}
\s_{\l,\m}(A)=
\sum_{n=0}^k
(-1)^{\left[\frac{n+1}{2}\right]}\,
\frac{
\left(
\begin{array}{c}
\left[\frac{k}{2}\right]\\[4pt]
\left[\frac{2n+1-(-1)^{n+k}}{4}\right]
\end{array}
\right)
\left(
\begin{array}{c}
\left[\frac{k-1}{2}\right]+2\l\\[4pt]
\left[\frac{2n+1+(-1)^{n+k}}{4}\right]
\end{array}
\right)
}
{
\left(
\begin{array}{c}
2(\m-\l)+n-k-1\\[4pt]
\left[\frac{n+1}{2}\right]
\end{array}
\right)
}\,
D^n(a)\,\a^{\m-\l+\frac{n-k}{2}}
\end{equation}
where $A=a(x,\xi)\,\overline{D}^k\in\cD^{\frac{k}{2}}_{\l,\m}$.

(ii)
In the resonant case the $\osp(1|2)$-modules $\cD_{\l,\m}$ and $\cS_{\m-\l}$
are not isomorphic, except if $(\l,\m)$ are given by (\ref{TheResVal}).
\end{thm}

Note that the binomial coefficients in (\ref{SymMapMainFor}) are defined by
${\nu\choose{}q}=\frac{\nu(\nu-1)\cdots(\nu-q+1)}{q!}$.
This expression makes sense for arbitrary $\nu\in\bbC$.

\begin{rmk}
{\rm
The map $\s_{\l,\m}$ is defined with the help of local coordinates $(x,\xi)$.
Nevertheless, this map is independent with respect to the fractional-linear
coordinate transformations (\ref{SuperLiFracForm}). In particular, if one
fixes a projective structure on $S^{1|1}$, the map $\s_{\l,\m}$ is globally
defined. This follows from the $\osp(1|2)$-equivariance.
}
\end{rmk}

Let us now give the explicit formula for the quantization map.

\begin{prop}
\label{MainThmbis}
The map $Q_{\l,\m}=\s^{-1}_{\l,\m}$ associates to a tensor density
$\varphi=f\,\a^{\m-\l-\frac{k}{2}}$ the following differential operator
from $\cF_\l$ to $\cF_\m$:
\begin{equation}
\label{QuantMapMainFor}
Q_{\l,\m}(\varphi)=
\sum_{n=0}^k
\frac{
\left(
\begin{array}{c}
\left[\frac{k}{2}\right]\\[4pt]
\left[\frac{2n+1-(-1)^{n+k}}{4}\right]
\end{array}
\right)
\left(
\begin{array}{c}
\left[\frac{k-1}{2}\right]+2\l\\[4pt]
\left[\frac{2n+1+(-1)^{n+k}}{4}\right]
\end{array}
\right)
}
{
\left(
\begin{array}{c}
2(\m-\l)-k+\left[\frac{n-1}{2}\right]\\[4pt]
\left[\frac{n+1}{2}\right]
\end{array}
\right)
}\,D^n(f)\,\overline{D}^{k-n}.
\end{equation}
\end{prop}

\begin{rmk}
{\rm
In \cite{CMZ}, the $\osp(1|2)$-quantization map was written down in
the particular case $\l=\m=0$. The authors use the
form (\ref{DiffOpGenFormEquiv}) of a differential operator.
One can rewrite formula~(\ref{QuantMapMainFor}) using the same notations:
$$
Q_{\l,\m}(f\,\a^{\m-\l-\frac{k}{2}})= \sum_{n\geq0} \frac{ \left(
\begin{array}{c}
\left[\frac{k}{2}\right]\\[4pt]
\left[\frac{n+1}{2}\right]
\end{array}
\right)
\left(
\begin{array}{c}
\left[\frac{k-1}{2}\right]+2\l\\[4pt]
\left[\frac{n}{2}\right]
\end{array}
\right)
}
{
\left(
\begin{array}{c}
k+2(\l-\m)\\[4pt]
\left[\frac{n+1}{2}\right]
\end{array}
\right)
}\,D^n(f)\,D^{k-n},
$$
where the upper bound for summation is
$2\left[\frac{k+1}{2}\right]$. In the particular case $\l=\m=0$,
this formula coincides with formula (7.5) of \cite{CMZ}. }
\end{rmk}

\subsection{Proof of Theorem \ref{MainThm}}

The proof of Theorem \ref{MainThm} consists of three parts. First,
we show that the symbol map (\ref{SymMapMainFor}) is, indeed,
$\osp(1|2)$-equivariant. We also prove the existence for the
special values (\ref{TheResVal}). Then we prove the uniqueness of
the symbol map (\ref{SymMapMainFor}). Finally, we show that, for
the resonant values of $\m-\l$, there is no isomorphism between
$\cD_{\l,\m}$ and $\cS_{\m-\l}$ if $\l$ and $\m$ are not given by
(\ref{TheResVal}).

\medskip
\textbf{A. Existence and explicit formula}.
Consider symbol maps given by differential operators of the form
\begin{equation}
\label{SymMap}
\s(A)= \sum_{n=0}^k \b_{n}^k(x,\xi)\, D^n(a)\, \a^{\m-\l+\frac{n-k}{2}},
\end{equation}
where $A=a(x,\xi)\,\overline{D}^k$, and
$\b_{n}^k(x,\xi)\in C_\bbC^\infty(S^{1|1})$ are arbitrary functions,
and calculate the condition of $\osp(1|2)$-equivariance. Clearly,
it suffices to impose invariance with respect to $D$ and $xD$ and the
following is straightforward.

\begin{lem}
\label{DEqivLem}
(i)
A symbol map (\ref{SymMap}) commutes with the action of $D$ if and only if
the coefficients $\b_{n}^k(x,\xi)$ are constants (i.e., do not depend
on $x,\xi$ and on the parity of $A$).

(ii)
A symbol map (\ref{SymMap}) commutes with the action of $xD$ if and only if
the following system is satisfied:
\begin{equation}
\label{Eq1}
\begin{array}{rcl}
p\b_{2m-1}^{2s-1}
&=&
m\b_{2m}^{2s}, \\[6pt]
s\b_{2m}^{2s-1}
&=&
(2(\l-\m)-m+2s)\,\b_{2m+1}^{2s}, \\[6pt]
(2\l+s)\,\b_{2m-1}^{2s}
&=&
m\b_{2m}^{2s+1}, \\[6pt]
(2\l+s)\,\b_{2m}^{2s}
&=&
(2(\l-\m)-m+2s+1)\,\b_{2m+1}^{2s+1},
\end{array}
\end{equation}
where $1\leq{}s\leq\left[\frac{k}{2}\right]$ and $0\leq{}m\leq{}s$.
\end{lem}

If $\m-\l$ is non-resonant, then it is easy to see that the solution of the
system (\ref{Eq1}) with the initial condition $\b_0^k=1$ (which is equivalent
to the fact that $\s$ preserves the principal symbol) is unique and
given by (\ref{SymMapMainFor}).

If $\m-\l$ is resonant and given by (\ref{TheResVal}), then
the system (\ref{Eq1}) can also be easily solved.
The solution is no longer unique since the system is split into separate
independent parts. Note that in this case, the explicit solution can still
be obtained from formula (\ref{SymMapMainFor})
if one chooses an arbitrary resolution of $0/0$-singularities.
\medskip

\textbf{B. Uniqueness}. Consider the non-resonant case. The main
ingredient of our proof of the uniqueness of an
$\osp(1|2)$-equivariant symbol map is the \textit{locality}
property and Definition 1 of differential operators, see Section
\ref{DDOSect}. This idea is borrowed from \cite{LO}. Since any
$\Sl(2)$-equivariant map
\begin{equation}
\label{DecoMap}
T:\cF_{\nu_1}(S^1)\to\cF_{\nu_2}(S^1),
\end{equation}
with $\nu_1\geq\nu_2$, is local (see Theorem 5.1 from \cite{LO}),
so is the symbol map. Indeed, given two $\osp(1|2)$-equivariant
symbol maps $\s_1$ and $\s_2$, the map
$$
\s_2^{-1}\circ\s_1:\cS_{\m-\l}\to\cS_{\m-\l}
$$
is $\osp(1|2)$-equivariant.
Decomposition (\ref{SpaceSymForm}) shows that $\s_2^{-1}\circ\s_1$ is a sum
of maps (\ref{DecoMap}). Therefore, $\s_2^{-1}\circ\s_1$ is a
differential operator. It follows that the general form of a symbol map
$\s$ commuting with the action of $\osp(1|2)$ is given by
formula (\ref{SymMap}). However, we already proved in part \textbf{A} that
such a symbol map is unique and given by (\ref{SymMapMainFor}).
\medskip

\textbf{C. Cohomological obstructions}. Assume now that the shift
of the weight is resonant, $\m-\l=\frac{m}{2}$, but $(\l,\m)$ are
not as in (\ref{TheResVal}). We will prove that there is no
$\osp(1|2)$-isomorphism between $\cD_{\l,\m}$ and $\cS_{\m-\l}$ in
this case. Indeed, recall (\cite{Fuk}, Section 1.4.5) that any
exact sequence of $\fg$-modules
$$
\begin{CD}
0@>>>V@>i>>
W@>>>
U@>>>0,
\end{CD}
$$
defines an element in $H^1(\fg;\Hom(U,V))$: for any section
$\tau:U\to{}W$, we define a 1-cocycle on $\fg$ by
$$
c_\tau(X)\,(u)=i^{-1}((\rho_X\circ\tau-
(-1)^{p(\tau)p(\rho_X)}\,\tau\circ\rho_X)\,(u)),
$$
where $u\in{}U$ and $\rho_X$ is the action of $X\in\fg$; the
cohomology class $[c_\tau]$ of $c_\tau$ is independent of the
choice of $\tau$. The sequence is split if and only if $[c_\tau]=0$.

The exact sequence of $\osp(1|2)$-modules
\begin{equation}
\begin{CD}
\label{ExSecdiag}
0@>>>\cD^\frac{m-1}{2}_{\l,\m}@>i>>
\cD^\frac{m}{2}_{\l,\m}@>\s_{\rm pr}>>
\cF_{0}@>>>0,
\end{CD}
\end{equation}
where $i$ is the inclusion, defines an element of
$H^1(\osp(1|2);\Hom(\cF_{0},\cD^\frac{m-1}{2}_{\l,\m}))$; the
principal symbol map $\s_{\rm
pr}:\cD^\frac{m-1}{2}_{\l,\m}\to\cF_\frac{1}{2}$ defines
$\bar{c}\in H^1(\osp(1|2);\Hom(\cF_0,\cF_\frac{1}{2}))$. Set:
$\tau:a\mapsto{}a\,\overline{D}^{m}. $ The following lemma is a
straightforward computation.

\begin{lem}
\label{VeryLastLem}
If $m$ is odd, then $\bar{c}=\left(\l+\frac{m-1}{4}\right)\g_1$;
if $m$ is even, then $\bar{c}=\frac{m}{2}\,\g_1$.
\end{lem}
\noindent If $\bar{c}\not=0$, then the sequence (\ref{ExSecdiag})
is not split. The module $\cD^\frac{m}{2}_{\l,\m}$ is then
non-isomorphic to the corresponding graded module of symbols.
Theorem \ref{MainThm} is proved. \qed

Proof of Proposition \ref{MainThmbis} is similar to that of
Theorem \ref{MainThm}, part (i).

\subsection{Discussion}

It is quite clear that the $\cK(1)$-modules $\cD^\ell_{\l,\m}$ and
$\cS^\ell_{\m-\l}$ are not isomorphic for $\ell>2$ and there is no
$\cK(1)$-invariant symbol map. For instance, one can check that
the unique $\osp(1|2)$-invariant symbol map from Theorem
\ref{MainThm} does not commute with any other element of $\cK(1)$,
implying the above statement in the non-resonant case.
However, we do not give here a complete proof.
Note that a similar result holds in the case of $S^1$, see  
\cite{Gar,GO1} and the
multi-dimensional case \cite{LO}. This problem is related to the first
cohomology space $H^1(\cK(1);\cD^\ell_{\l,\m})$ (cf. \cite{BO}
for preliminary results on this subject).

We believe that the $\cK(1)$-modules $\cD^\ell_{\l,\m}$ deserve further study.
We formulate here the problem of classification of these modules
(see \cite{Gar,GO1} for the case of $S^1$), as well as existence of the
exceptional weights $(\l,\m)$ in the sense of \cite{Con}.
It would also be interesting to study the corresponding automorphism
groups (see \cite{GMO}).

\vskip 0.2cm

\noindent \textbf{Acknowledgments}.
We are grateful to C.~Conley and D.~Leites for very careful
readings of this paper at various stages of elaboration, valuable comments
and correction of a number of mistakes.
We are pleased to thank C.~Duval and P.~Lecomte for their interest
in this work and enlightening discussions.

\end{document}